\documentclass[10pt,a4paper,english,nofootinbib,twocolumn]{revtex4}
\usepackage{lmodern}

\usepackage[T1]{fontenc}
\usepackage[latin9]{inputenc}
\usepackage{babel}
\usepackage{amsmath}
\usepackage{graphicx}
\usepackage{amssymb}
\usepackage{esint}
\usepackage[unicode=true, pdfusetitle,
bookmarks=true,bookmarksnumbered=false,bookmarksopen=false,
breaklinks=false,pdfborder={0 0 1},backref=false,colorlinks=false]
{hyperref}
\def\b{\begin{equation}}
\def\e{\end{equation}}

\makeatletter
\@ifundefined{textcolor}{}
{%
 \definecolor{BLACK}{gray}{0}
 \definecolor{WHITE}{gray}{1}
 \definecolor{RED}{rgb}{1,0,0}
 \definecolor{GREEN}{rgb}{0,1,0}
 \definecolor{BLUE}{rgb}{0,0,1}
 \definecolor{CYAN}{cmyk}{1,0,0,0}
 \definecolor{MAGENTA}{cmyk}{0,1,0,0}
 \definecolor{YELLOW}{cmyk}{0,0,1,0}
 }

\usepackage{latexsym}\usepackage{bm}

\makeatother

\makeatother

\begin{document}

\title{Graviton Mass and Memory}

\author{Ercan Kilicarslan}

\email{ercan.kilicarslan@usak.edu.tr}

\affiliation{Department of Physics,\\
 Usak University, 64200, Usak, Turkey}

\author{Bayram Tekin}

\email{btekin@metu.edu.tr}

\affiliation{Department of Physics,\\
 Middle East Technical University, 06800, Ankara, Turkey}
\date{\today}

\begin{abstract}
Gravitational memory, a residual change, arises after a finite gravitational wave pulse interacts with free masses.  We calculate the memory effect in massive gravity as a function of the graviton mass $(m_g)$ and show that it is discretely different from the result of general relativity: the memory is reduced not just via the usual expected Yukawa decay but by a numerical factor which survives even in the massless limit. For the strongest existing bounds on the graviton mass, the memory is essentially wiped out for the sources located at distances above 10 Mpc. On the other hand, for the weaker bounds found in the LIGO observations, the memory is reduced to zero for distances above 0.1 Pc. Hence, we suggest that careful observations of the gravitational wave memory effect can rule out the graviton mass or significantly bound it. We also show that adding higher curvature terms reduces the memory effect.

\end{abstract}

\maketitle

\section{Introduction}

There is a very natural question that one can ask about the gravitational
waves that have been detected by the LIGO/VIRGO:  did the  detectors
leave a permanent effect on the waves or the waves left the detectors
intact as they entered ? Of course, one can formulate the problem in just
the opposite way: did the waves leave a permanent effect on the detectors ? The  second
formulation is better because we might be able to measure the effect if
that is the case. It turns out, for certain gravitational waves, part of
the strain can be considered as a sort of permanent effect on the
detector. This phenomenon is aptly called the {\it gravitational memory effect} and comes in two
related forms: ordinary (or linear) \cite{zeldovich},  null  (or
nonlinear) \cite{Christodoulou} which could be measured soon in the observations.

One might wonder why this somewhat subtle effect arises in the first place. Let us explain this a little bit. 
All the effects of gravity are encoded in the metric tensor field $g$ which needs no coordinates 
to be defined. General Relativity (GR) is intrinsically four dimensional and the full metric
of {\it spacetime} manifold $g$ does not really evolve in time: it is
what it is. So, if we had known how to obtain all the local observables  from the metric for all physically relevant situations, we would not need any further nomenclatures such as memory effect, gravitomagnetism {\it etc.} But, since  as local observers, we do not have full access to the fully consistent spacetime, it pays to see spacetime as space evolving in time, namely, to see spacetime as a history of space. Such a dynamical picture requires a choice of time and other coordinates and leads to interesting  phenomena and the gravitational memory is one such an event: the wave that enters the interaction with the detector masses  differs in some well-defined sense  from the wave that leaves the interaction. The best way to see the difference is to measure the change in the relative separation of the masses as this is related to the change in the wave profile. In geometric units, in GR, the total change of the wave profile is given by two parts
\begin{equation}
\Delta h^{\textnormal{TT}}_{ab} = \Delta {h_1}^{\textnormal{TT}}_{ab} +\Delta {h_2}^{\textnormal{TT}}_{ab},
\label{firsteq}
\end{equation}
where the first part comes from the massive unbound sources with masses $m_i$ and velocities $v_i$ and is given as \cite{Braginsky:1985fw}
\begin{equation}
\Delta {h_1}^{\textnormal{TT}}_{ab}= \frac{4}{r} \Delta\sum_i \frac{m _i}{\sqrt{1-v_i^2}}\left[\frac{(v_i)_a (v_i)_b}{1-v_i\cos\theta_i}\right]^{\textnormal{TT}}. 
\label{nonrev1}
\end{equation}
Here the unbound sources are located at the origin and $r$ is the radial coordinate of the detector located at a distance far away from the sources. $\Delta$ before the summation denotes the difference after and before the wave interacts with the detector, the TT index refers to the transverse-transpose component while the indices $a$ and $b$  are abstract spacetime indices \cite{R.M.Wald}. $\theta_i$ is the angle between the velocity  $v_i$ and $\hat{r}$. The second part  in (\ref{firsteq}) is somewhat more subtle and was initially found by Christodoulou \cite{Christodoulou} as a result of carefully studying the change at the null infinity once a null stress energy-tensor reaches the null infinity. A more transparent physical interpretation was given by Thorne \cite{Thorne:1992sdb}: considering each graviton emitted by the source as an unbound system, one should simply modify (\ref{nonrev1}) to take into account the gravitons as 
\begin{equation}
\Delta {h_2}^{\textnormal{TT}}_{ab}= \frac{4}{r}\Delta\int \frac{dE}{d\Omega'} \left[\frac{\zeta'_{a} \zeta'_b}{1-\cos\theta'}\right]^{\textnormal{TT}}d\Omega' \, ,
\label{nonrev}
\end{equation}
where $\Omega'$ is the solid angle, $\zeta'$ is the unit vector in the direction of the solid angle, $\theta'$ is the angle between the unit vector $\hat{r}$ and $\frac{dE}{d\Omega'}$ describes the radiated energy that reaches the null infinity per unit angle.

In this work, we calculate the gravitational memory as a function of the graviton mass and suggest that a possible observation of memory can constrain or possibly rule out the graviton mass. We shall also discuss the memory effect in quadratic gravity. A priori one would expect that the effect of having a massive graviton (with mass $m_g$) amounts to a change of the physically relevant quantities such as the $\frac{1}{r}$ potential to $\frac{e^{-m_gr}}{r}$, which indeed is true but the overall factor is not correct: the weak field limit of the Newtonian potential in the massive gravity is $V(r)=-\frac{4G}{3}\frac{e^{-m_gr}}{r}$, which has the well-known Van-dam-Veltmann-Zakharov (vDVZ) \cite{vdvz1,vdvz2} discontinuity that cannot be remedied by redefining the Newton's constant as that would lead to a wrong prediction of light deflection by the Sun. Relativistic counterparts of the vDVZ discontinuity have been found recently \cite{Gullu-Tekin, Tasseten-Tekin} where massive gravity predicts a maximized total spin for two interacting bodies, while Einstein's gravity predicts a minimum total spin. Here we study the effects of graviton mass and quadratic terms on gravitational memory and show that it is significantly different from that of GR in the case of massive gravity. 

The lay out of the paper is as follows: In section II, we calculate the memory effect in the low energy massive gravity theory (namely the Fierz-Pauli theory). The computation boils down to solving the geodesic deviation equation in the presence of the Riemann tensor which is determined by a passing gravitational wave in massive gravity. In section III, we carry out a similar calculation for quadratic gravity, that has a massive spin-0 and a massive spin-2 particle along with the Einsteinian massless spin-2 particle. The computation is in generic $D$ dimensions. In the Appendix, we consider the massive scalar field case to set the notation and our conventions, especially how we define the sources that create the fields. 

\section{Memory effect in massive gravity}
The action for massive gravity is 
\begin{equation}
 {I}=\int d^4x \sqrt{-g} \, \Big (  \frac{1}{ \kappa}R -\frac{m_g^2}{4 \kappa}(h^2_{ab}-h^2)+{\cal{L}}_{matter}\bigg),
\label{pfaction}
\end{equation}
which yields the linearized field equations 
\begin{equation}
{\cal G}^L_{ab}+\frac{m_g^2}{2}(h_{ab}-\bar{g}_{ab}h)=8\pi T_{ab},
\label{PFeom}
\end{equation}
where ${\cal G}^L_{ab}$ is the linearized Einstein tensor and $\bar{g}_{ab}$ refers to the background metric  (see \cite{deser_tekin} for the relevant definitions of linearized tensors). Assuming a flat background ($\bar{g}_{ab} = \eta_{a b}$) and a conserved source ( $\partial_a T^ {a b} =0$), one arrives at 
\begin{equation}
\begin{aligned}
(\partial^2-m_g^2)h_{ab}=&-16\pi\bigg(T_{ab}-\frac{1}{3}(\eta_{ab}-\frac{1}{m_g^2}\partial_a\partial_b)T\bigg)\\&\equiv-16\pi\tilde{T}_{ab},
\end{aligned}
\end{equation}
whose inhomogeneous solution can be written as  
\begin{equation}
h_{ab} = 16\pi\int G_{ab}{}^{cd}(x,x')\tilde{T}_{cd}(x')d^{4}x',
\label{Solution}
\end{equation}
with the retarded Green's function given as 
\begin{equation}
G_{ab}{}^{cd}(x,x')=\eta_{a}{}^{c}\eta_{b}{}^{d}G(x,x'),
\label{Green1}
\end{equation}
here $\eta_{a}{}^{c}$ is the parallel propagator. We follow the analogous computation in GR \cite{Satishchandran:2017pek, Garfinkle}, see the Appendix below for the case of 
the massive scalar field where we establish the notation. Now consider the source to be some free particles colliding at the point $t=0, \vec{x}$ and some (possibly other) particles coming out from that single spacetime point. Then the energy momentum tensor of the source is 
\begin{equation}
\begin{aligned}
T_{ab}=&\sum_{(j)in}m^{\rm in}_{(j)}\frac{d\tau_{(j)}}{dt}u_{(j)a}u_{(j)b}\delta_{3}(\mathbf{x}-\mathbf{y}_{(j)}(t))\Theta(-t) \\&+ \sum_{(i)out}m^{\rm out}_{(i)}\frac{d\tau_{(i)}}{dt}u_{(i)a}u_{(i)b}\delta_{3}(\mathbf{x}-\mathbf{y}_{(i)}(t))\Theta(t),
\label{SourceSol}
\end{aligned}
\end{equation}
where $u_{(i)a}$ and $u_{(j)a}$ are normalized four velocities and the propagator can be given as
 \begin{equation}
  G^R(x,x')=(\partial^2-m_g^2)^{-1}=\frac{1}{4\pi r}e^{-m_gr}\delta(t-t'-r)
  \label{GreenSol}.
 \end{equation}
Using (\ref{GreenSol}), (\ref{SourceSol}), (\ref{Green1}) in (\ref{Solution}), the retarded solution for massive gravity theory can be obtained as
\begin{equation}
\begin{aligned}
  h_{ab}(x)=&\bigg[4 \bigg(\alpha_{ab}\Theta(U)+\beta_{ab}\Theta(-U)\bigg)+\frac{4}{3m_g^2}\bar{g}_{cd} \\&\partial_a  \partial_b\bigg(\tilde{\alpha}^{cd}\Theta(U)+\tilde{\beta}^{cd}\Theta(-U)\bigg)\bigg]\frac{e^{-m_gr}}{r},
  \label{Field1}
  \end{aligned}
 \end{equation}
where $U  \equiv t -r$ is the retarded time and  we defined 
\begin{equation}
\begin{aligned}
&\alpha_{ab}(\hat{\mathbf{r}}) \equiv\sum_{(i)out}\frac{d\tau^{(i)}}{dt}\Big(\frac{m^{\rm out}_{(i)}}{1-\hat{\mathbf{r}}\cdot\mathbf{v}^{(i)}}\Big)\bigg(u_{a}^{(i)}u_{b}^{(i)}+\frac{1}{3}\eta_{ab}\bigg), \\
&\tilde{\alpha}_{ab}(\hat{\mathbf{r}}) \equiv \sum_{(i)out}\frac{d\tau^{(i)}}{dt}\Big(\frac{m^{\rm out}_{(i)}}{1-\hat{\mathbf{r}}\cdot\mathbf{v}^{(i)}}\Big)u_{a}^{(i)}u_{b}^{(i)}. 
\end{aligned}
\end{equation}
We did not write the explicit form of $\beta_{ab}$ since it is exactly like $\alpha_{ab}$, except  one replaces $\text{``out``}$ with $\text{``in``}$ which is also the case for  $\tilde\beta_{ab}$.  Already at this stage there seems to be two differences between massive gravity and the massless GR: due to the second term in (\ref{Field1}),
the $m_g\rightarrow 0$ limit seems divergent, but this is a red-herring, that term does not contribute to the linearized Riemann tensor and so it is of no real consequence. But in $\alpha_{ab}(\hat{\mathbf{r}})$ the factor  1/3 in front of $\eta_{a b}$ is 1/2 in massless GR. This will be crucial as the rest smoothly reproduces the GR result in the massless limit. For the moment keeping all the terms in  (\ref{Field1}), one can find up to the leading order in $\frac{1}{r}$
\begin{equation}
\begin{aligned}
  h_{ab}(x)=&4\bigg(\alpha_{ab}\Theta(U)+\beta_{ab}\Theta(-U)\bigg)\frac{e^{-m_gr}}{r}\\&+\frac{4}{ 3\, m_g^2}\bar{g}_{cd}\bigg((\tilde{\alpha}^{cd}\Theta(U)+\tilde{\beta}^{cd}\Theta(-U))(m_g^2r_ar_b)\\&+m_g(\tilde{\alpha}^{cd}-\tilde{\beta}^{cd})(K_ar_b+K_br_a)\delta(U)\\&+(\tilde{\alpha}^{cd}-\tilde{\beta}^{cd})K_aK_b\delta'(U)\bigg)\frac{e^{-m_gr}}{r},
  \label{field}
  \end{aligned}
 \end{equation}
where $K^a \equiv -\partial^aU=t^a+r^a$ and $t^a$ and $r^a=\partial^ar$ are unit vectors. We can now compute the linearized Riemann tensor which is
\begin{equation}
R_{abcd}= \partial_{c}\partial_{[b}h_{a]d} - \partial_{d}\partial_{[b}h_{a]c}.
\end{equation}
Note that up to ${\cal{O}}(\frac{1}{r^2})$, one has 
\begin{equation}
\begin{aligned}
\partial_d\partial_a\bigg(\frac{e^{-m_gr}}{r}\Theta(U)\bigg)=&\bigg(m_g^2r_ar_d\Theta(U)\\&+m_g\delta(U)(K_ar_d+K_dr_a)\\&+\delta'(U)K_aK_d\bigg)\frac{e^{-m_gr}}{r}.
\end{aligned}
\end{equation}
As noted above, the $1/m_g^2$ terms in  (\ref{Field1}) do not contribute to Riemann tensor. Finally, to the leading order, the linearized Riemann tensor reads
\begin{equation}
\begin{aligned}
R_{abcd}=&4\bigg(K_{[a}\Delta_{b][c}K_{d]}\frac{d^2\Theta(U)}{dU^2}+m_gK_{[a}\Delta_{b][c}r_{d]}\frac{d\Theta(U)}{dU}\\&+m_gK_{[d}\Delta_{b][c}r_{a]}\frac{d\Theta(U)}{dU}+2m_g^2r_{[a}\alpha_{b][c}r_{d]}\Theta(U)\\&+2m_g^2r_{[a}\beta_{b][c}r_{d]}\Theta(-U)\bigg)\frac{e^{-m_gr}}{r},
 \label{Riemann}
 \end{aligned}
\end{equation}
where $\Delta_{ab} \equiv 2  (\alpha_{ab}(\hat{\mathbf{r}}) -\beta_{ab}(\hat{\mathbf{r}}))$.
In the GR case, the linearized Riemann tensor is gauge invariant and one can work in any gauge one likes and the TT gauge is the most convenient one, hence the TT indices in all the expressions. But in massive GR, as the symmetry of the theory, one only has the rigid background symmetries, not the the full linearized diffeomorphisms, one cannot consider the TT gauge. The memory part of the linearized Riemann tensor is only the first term in (\ref{Riemann}) as can be seen from the computation of the geodesic deviation between two massive sources at rest with a relative separation vector $\xi$:
\begin{equation}
\frac{d^{2}\xi^{i}}{dt^{2}}=-{R^i}_{0j0}\xi^{j}.
\label{geodesicdevmemory}
\end{equation}
Plugging (\ref{Riemann}) into the last equation and integrating twice yield
\begin{equation}
\begin{aligned}
\Delta\xi^{i}&=\int_{-\infty} ^{U}dU'\int_{-\infty} ^{U'}dU''\frac{d^{2}\xi^{i}}{dU''^{2}}\\
&=\frac{1}{r}e^{-m_gr}\tilde{\Delta}_j^i(m_g)\Theta(U)\xi^{j},
\label{memoryeffect}
\end{aligned}
\end{equation} 
where the memory tensor $\tilde{\Delta}_j^i(m_g)$ is given explicitly as
\begin{equation}
\tilde{\Delta}_j^i(m_g)\equiv \Delta_j^i(m_g)+\delta_j^i \Delta_{00}(m_g)+\hat{r}^i \Delta_{0j}(m_g)+\hat{r}_j \Delta_0\,^i(m_g).
\end{equation} 
In GR, one only has the first part and moreover the relation between the memory tensors in massive gravity and GR is 
\begin{equation}
\Delta_{ a b}(m_g) = \Delta_{a b}(\text{GR}) - \frac{1}{6}\eta_{a b} \eta^{ c d}\Delta_{ cd } (\text{GR}). 
\label{mem}
\end{equation}
Let us give some numerical values: as the graviton mass is expected to be small ($ m_g < 10^{-29 }$ eV $\approx 5\times 10^{-20} \frac{1}{\text{km}}$) \cite{nieto}, for small $r$,  we can take $m_g r \rightarrow 0$ and  the Yukawa decay part reduces to the usual Einsteinian $1/r$ form, but the noted discrete difference survives and an accurate measurement of memory can distinguish massive gravity from GR as  $\Delta_{ a b}(m_g \rightarrow 0) \ne \Delta_{a b}(\text{GR})$. On the other hand,  if $r = 1$ Mpc, then one has $m_g r \approx 1.55$ and the memory is reduced by $0.21$ due to the Yukawa decay part. For larger separations, as in the case of the first black hole merger observation which was at a distance $440^{+160}_{-180}$ Mpc \cite{ligo}, all the memory is wiped out in massive gravity. For weaker bounds on the graviton mass, such as the one noted in \cite{ligo2} ($ m_g < 7.7 \times 10^{-23}$ eV), the memory is wiped out virtually above 0.1 Pc !

\section{Higher Derivative Gravity}
In \cite{Garfinkle}, the authors showed that there is no gravitational memory effect in higher even dimensional  spacetimes ($D>4$). Here we add quadratic curvature terms (which are the only relevant ones in flat backgrounds in the weak-field limit)  to the Einstein's theory and compute the memory effect in generic $D$ dimensions:
\begin{equation}
\begin{aligned}
I  = \int d^{{D}}x\,\sqrt{-g}\{&
\frac{1}{\kappa}R+\alpha R^{2}+\beta
R_{ab}^{^{2}}
+\gamma(R_{abcd}^{2}\\&-4R_{ab}^{2}+R^{2})+ {\cal {L}}_{\mbox{matter}}\} ,
\label{action11}
\end{aligned}
\end{equation}
which yields the linearized field equations around the flat background metric 
\begin{equation}
\begin{aligned}
 \frac{1}{\kappa} {\mathcal{G}}_{ab}^{L}
  +  \left(2\alpha+\beta\right)\left(\bar{g}_{ab}\partial^2-\partial_{a}\partial_{b}
 \right)R^{L}
 +  \beta\partial^2{\mathcal{G}}_{ab}^{L}=T_{ab}\left(h\right).
\label{linearized11}
\end{aligned}
\end{equation}
In the harmonic gauge $\partial^{a}h_{ab}=\frac{1}{2} \partial_{b}h $, the linearized field equations reduce to
\begin{equation}
\begin{aligned}
 (\frac{1}{\kappa}+ \beta \partial^2) \partial^2  h_{ab}=& -2T_{ab}+ 2(2\alpha+\beta)(\bar{g}_{ab}\partial^2-\partial_a \partial_b)R^L\\&
 - (\frac{1}{\kappa}+\beta \partial^2) \bar{g}_{ab}R^L,
 \end{aligned}
\end{equation}
whose inhomogeneous solution reads
\begin{equation}
\begin{aligned}
h_{ab} = \int & d^{D}x'\bigg(2 G^1(x,x')T_{ab}(x')+2 \bar{g}_{ab}G^3(x,x')T(x')\\&-4(2\alpha+\beta)G^2(x,x')(\bar{g}_{ab}\partial^2-\partial_a \partial_b)T(x')\bigg),
 \end{aligned}
\end{equation}
with the retarded scalar Green's functions given as 
\begin{equation}
\begin{aligned}
&G^1(x,x')=\frac{1}{\beta}\bigg(( \partial^2-m_\beta^2) \partial^2 \bigg)^{-1},\\&G^2(x,x')=\frac{\bigg((\partial^2-m_\beta^2) (\partial^2 - m_c^2)\partial^2\bigg)^{-1}}{\beta\left( 4 \alpha (D-1) + D\beta \right)}\\& G^3(x,x')=\frac{1}{\left( 4 \alpha (D-1) + D\beta \right)}\bigg((\partial^2 - m_c^2)\partial^2\bigg)^{-1}.
\end{aligned}
\end{equation}
Here the mass of the massive spin-$2$ and the massive spin-$0$ graviton are given as $m_\beta^2=-\frac{1}{\beta\kappa}$, $m_c^2=\frac{D-2}{\kappa\left( 4 \alpha (D-1) + D\beta \right)}$, respectively. 
By using these, to leading order after a somewhat cumbersome calculation, the linearized Riemann tensor can be found as
\begin{equation}
\begin{aligned}
R_{abcd}=&\frac{\kappa}{(2\pi r)^{\frac{D-2}{2}}}K_{[a}\bar{\Delta}_{b][c}K_{d]}\frac{d^{\frac{D-2}{2}}}{dU^{\frac{D-2}{2}}}\delta(U)\\&-\frac{\kappa e^{-m_\beta r}}{(2\pi r)^{\frac{D-2}{2}}}(m_\beta)^{\frac{D-4}{2}}\bigg(K_{[a}\bar{\Delta}_{b][c}K_{d]}\delta'(U)\\&+m_\beta K_{[a}\bar{\Delta}_{b][c}r_{d]}\delta(U)+m_\beta K_{[d}\bar{\Delta}_{b][c}r_{a]}\delta(U)\\&+2m_\beta^2r_{[a}\bar{\alpha}_{b][c}r_{d]}\Theta(U)+2m_\beta^2r_{[a}\bar{\beta}_{b][c}r_{d]}\Theta(-U)\bigg),
 \label{Riemann1}
 \end{aligned}
\end{equation}
here we have defined 
\begin{equation}
\begin{aligned}
&\bar{\Delta}_{ab} \equiv 2\sum_{(i)out} \frac{d\tau_{(i)}}{dt}\Big(\frac{m^{\rm out}_{(i)}}{1-\hat{\mathbf{r}}\cdot\mathbf{v}_{(i)}}\Big)\bigg(q_{ac}u^{c}_{(i)}q_{bd}u^{d}_{(i)}\\
&-\frac{q_{cd}u^{c}_{(i)}u^{d}_{(i)}}{D-2}q_{ab}\bigg) -2\sum_{(j)in}\frac{d\tau_{(j)}}{dt}\Big(\frac{m^{\rm in}_{(j)}}{1-\hat{\mathbf{r}}\cdot\mathbf{v}_{(j)}}\Big)\times\\
&\bigg(q_{ac}u^{c}_{(j)}q_{bd}u^{d}_{(j)}-\frac{q_{cd}u^{c}_{(j)}u^{d}_{(j)}}{D-2}q_{ab}\bigg),\\&
\bar{\alpha}_{ab}= \sum_{(i)out} \frac{d\tau_{(i)}}{dt}\Big(\frac{m^{\rm out}_{(i)}}{1-\hat{\mathbf{r}}\cdot\mathbf{v}_{(i)}}\Big)\bigg(q_{ac}u^{c}_{(i)}q_{bd}u^{d}_{(i)}\\
&-\frac{q_{cd}u^{c}_{(i)}u^{d}_{(i)}}{D-2}q_{ab}\bigg),
\label{memorytensor1}
\end{aligned}
\end{equation}
where $q_{ab}$ is the projector that projects a symmetric tensor onto the sphere $S^{D-2}$ at large $r$ and $\bar{\beta}_{ab}$ is exactly like $\bar{\alpha}_{ab}$, except  one replaces $\text{``out``}$ with $\text{``in``}$.
By using (\ref{geodesicdevmemory}), the finite relative change in the displacement between two free test particles can be computed as
\begin{equation}
\begin{aligned}
\Delta\xi^{i}=\frac{2\pi}{(2\pi r)^{\frac{D-2}{2}}}\bigg(\frac{d^{\frac{D-4}{2}}}{dU^{\frac{D-4}{2}}}- (m_\beta)^{\frac{D-4}{2}}e^{-m_\beta r}\bigg)\bar{\Delta}_j^i\Theta(U)\xi^{j},
\label{memoryeffect11}
\end{aligned}
\end{equation} 
here $\bar{\Delta}_j^i$ are spatial components of the memory tensor Eq.(\ref{memorytensor1}).
Observe that, in higher dimensional even spacetimes ($D> 4$), to the leading order, there is no memory effect as in the case of pure GR.  On the other hand, in four dimensions, the memory effect is 
\begin{equation}
\begin{aligned}
\Delta\xi^{i}=\frac{1}{r}\bigg(1- e^{-m_\beta r}\bigg)\bar{\Delta}_j^i\Theta(U)\xi^{j}.
\end{aligned}
\end{equation} 
In the $m_\beta \to \infty$, the memory is the same as obtained in \cite{Garfinkle}. But for any finite value of $m_\beta$, the memory is reduced compared to GR.
\section{Conclusions}
Recently gravitational memory effect received a renewed interest  \cite{Pasterski,Strominger1,Strominger2,Flanagan,Garfinkle,Hollands,Satishchandran:2017pek,Tolish1,Tolish2,Bieri2,Zhang,Kilicarslan1} for various reasons some of which are: its related to black hole soft hair, asymptotic symmetries and its potential observation in the gravitational wave detectors. Here, we calculated the gravitational memory as a function of graviton mass and showed that for the graviton mass $m_g \le 10^{-29}$ eV, the memory is significantly reduced for distances beyond $1$ Mpc as in the first observation of two black hole mergers which was at a distance of more than $200$ Mpc. Moreover massive gravity leaves a discretely different memory on our detectors from the expected general relativity result. The result is summarized by equation (\ref{mem}). In the LIGO/VIRGO observations of gravitational waves, memory effect is already in the data but it is hard to distinguish it is from the background noise. In the near future, one might expect to see this effect observed (possibly in eLISA). This observation might rule out massive gravity. We have also calculated the memory effect in quadratic gravity and showed that due to the massive spin-2 mode, the memory is reduced from that of the Einstein's theory. Here we have used the linearized massive gravity theory which is valid in the weak-field regime that is relevant for the gravitational wave bursts observed on earth. Of course one can consider non-linear extensions of massive gravity such as the one given in \cite{deRham:2010kj} but, the above result is universal in the weak field limit as the non-linear extensions reduce down to the Einstein-Fierz-Pauli theory that we employed.

\section{Appendix}

We follow the analogous computation in GR \cite{Garfinkle,Satishchandran:2017pek} and first establish the relevant Green's function for the scalar field case: consider a scalar source $S$ coupled to massive wave field in a $4$-dimensional Minkowski spacetime 
\begin{equation}
(\eta^{ab}\partial_{a}\partial_{b}-m^2)\phi = -4\pi S,
\label{waveeq1}
\end{equation}
from which follows the retarded Green's function 
\begin{equation}
G(x,x')=\frac{e^{-mr}}{4\pi r}\delta(t-t'-r),
\label{Green'sscalar1}
\end{equation}
yielding the general (inhomogeneous) solution of the Eq.(\ref{waveeq1}) as
\begin{equation}
\phi_S(x) = 4\pi\int{G(x,x')S(x')d^{4}x'}.
\label{phiintegralexp1}
\end{equation}
Now consider the source to be some free particles colliding at the point $t=0, \vec{x} =0$ and some (possibly other) particles coming out from that single spacetime point. Then the source is
\begin{equation}
\begin{aligned}
S(x)=&\sum_{(j)in}q^{\rm in}_{(j)}\frac{d\tau_{(j)}}{dt}\delta_{3}(\mathbf{x}-\mathbf{y}_{(j)}(t))\Theta(-t) \\&+ \sum_{(i)out}q^{\rm out}_{(i)}\frac{d\tau_{(i)}}{dt}\delta_{3}(\mathbf{x}-\mathbf{y}_{(i)}(t))\Theta(t),\label{gensource1}
\end{aligned}
\end{equation}
in which $q^{\rm out}_{i}$ $(q^{\rm in}_{j})$ are the out (in) scalar charges and $\tau_{(i)}$ is the proper time.
We would like to solve the equation \eqref{phiintegralexp1} for the source \eqref{gensource1}. For this purpose, for the sake of simplicity, let us consider a single created particle at $O$. The source can be given
\begin{equation}
S_0=q\delta_{3}(\mathbf{x})\Theta(t).
\label{createdscalar1}
\end{equation}
Plugging this into (\ref{phiintegralexp1}) and using the retarded Green's function (\ref{Green'sscalar1}), one gets
\begin{equation}
\phi_0(x)= q\int_0^\infty{\frac{1}{ r}e^{-mr}\delta(t-t'-r)dt'}.
\label{phi01}
\end{equation}
The solution reads 
\begin{equation}
\phi_0 = q\Theta(U)\frac{e^{-mr}}{r}.
\end{equation}
To obtain the field of a particle created at $O$ with the coordinate velocity $\mathbf{v}=d\mathbf{y}/dt$, Eq.(\ref{phi01}) can be boosted to get
\begin{equation}
\phi_{0,v}(x)=q\frac{d\tau}{dt}\bigg(\frac{1}{1-\hat{\mathbf{r}}\cdot\mathbf{v}}\bigg)\Theta(U)\frac{e^{-mr}}{r},
\label{phioutv}
\end{equation}
where  $\hat{\mathbf{r}}=\mathbf{x}/r$ is a unit vector.  Let us now consider the case that the particle is destroyed, the source is simply given
\begin{equation}
\tilde{S}_0=q\delta_{3}(\mathbf{x})\Theta(-t).\label{destroyedscalar}
\end{equation}
The solution is 
\begin{equation}
\tilde{\phi}_0 = q\Theta(-U)\frac{e^{-mr}}{r}.
\end{equation}
The linear superposition of the retarded  solutions for the case that the particles are created and destroyed can be written as
\begin{equation}
\phi_{S}(x)=(\alpha(\hat{\mathbf{r}})\Theta(U)+\beta(\hat{\mathbf{r}})\Theta(-U))\frac{e^{-mr}}{r},
\label{phiSv}
\end{equation}
where
\begin{equation}
\begin{aligned}
\alpha(\hat{\mathbf{r}})=\sum_{(i)out}q^{\rm out}_{(i)}\frac{d\tau_{(i)}}{dt}\bigg(\frac{1}{1-\hat{\mathbf{r}}\cdot\mathbf{v}_{(i)}}\bigg),
\end{aligned}
\label{alphabeta}
\end{equation}
and $\beta(\hat{\mathbf{r}})$ reads exactly the same except "out" becomes "in".

\end{document}